\documentclass[10pt,letterpaper]{article}
\usepackage{opex3}

\begin{document}

\title{Evolution of light trapped by a soliton in a microstructured
fiber}

\author{S. Hill, C. E. Kuklewicz, U. Leonhardt, F. K\"onig}

\address{School of Physics and Astronomy, University of St
Andrews, North Haugh, St Andrews, Fife, KY16 9SS, UK.}

\email{fewk@st-andrews.ac.uk} 
\homepage{http://www.st-andrews.ac.uk/~qinfo/}

\begin{abstract} We observe the dynamics of pulse trapping in a
microstructured fiber. Few-cycle pulses create a system of two
pulses: a Raman shifting soliton traps a pulse in the normal
dispersion regime. When the soliton approaches a wavelength of
zero group velocity dispersion the Raman shifting abruptly
terminates and the trapped pulse is released. In particular, the
trap is less than 4\,ps long and contains a 1\,ps pulse. After
being released, this pulse asymmetrically expands to more than
10\,ps. Additionally, there is no disturbance of the trapping
dynamics at high input pulse energies as the supercontinuum
develops further.
\end{abstract}

\ocis{(060.7140) Ultrafast processes in fibers; (190.5530) Pulse
propagation and temporal solitons; (190.4370) Nonlinear optics,
fibers}



\section{Introduction}
When a short and intense pulse of light is launched into a
microstructured fiber (MF), nonlinear effects can significantly
broaden the spectrum to much more than an octave
\cite{DudleyGenty06}. These supercontinua are used in a variety of
applications such as ultrafast optical switching, spectroscopy,
optical coherent tomography, optical clocks, etc. Routinely,
around 100-fs pulses are used at a wavelength close to a
zero-group velocity dispersion wavelength (ZDW) of the fiber.
Using nearly octave--spanning input pulses \cite{IshiiTeisset06,
SerebryannikovZheltikov05, FalkFrosz08}, the phenomenon of `pulse
trapping' \cite{NishizawaGoto02-1, NishizawaGoto02-2,
NishizawaGoto03, GorbachSkryabin07, TraversRulkov08,
CumberlandTravers08} can be observed. In this effect a
non-dispersing pulse can form at the short wavelength end of the
broad spectrum. This is surprising because the pulse exists in a
region of normal dispersion in the fiber, where the
dispersion-induced chirp cannot be cancelled by self-phase
modulation. The non-dispersing pulse is trapped behind a
fundamental soliton in the anomalous dispersion regime that was
generated by soliton fission. Soliton fission can create multiple
fundamental solitons, which shift to longer wavelengths via the
soliton self-frequency shift (SSFS) \cite{HusakouHerrmann01}.

Nishizawa and Goto \cite{NishizawaGoto02-2} were the first to
demonstrate that a soliton undergoing the SSFS can trap light
behind it. The trapped light adjusts to a wavelength that is group
velocity matched to the soliton. Because it is forced to travel
with the soliton and lies in the normal dispersion regime, it has
to shift to shorter wavelengths in order to keep the same
group-velocity as the soliton. Gorbach and Skryabin provided an
alternative view of pulse trapping \cite{GorbachSkryabin07}: the
light is trapped because the soliton accelerates. In simulations
they turned off the (negative) acceleration induced by the SSFS
and the trapping ceased. The acceleration of the soliton provides
a `gravity-like' potential to the trapped pulse.

In this paper we study experimentally the phenomenon of pulse
trapping in a fiber with two ZDWs. We use intense few-cycle pulses
to generate a soliton and a trapped pulse and focus in particular
on the trap dynamics as the soliton reaches the longer
zero-dispersion wavelength and decays. We observe how the light
escapes the trap, expanding to a few times the trap length. For
higher input pulse energies, a further component in the spectrum
at even shorter wavelengths is formed, which dominates the blue
end of the evolving supercontinuum.

\section{Pulse Trapping}

As previously shown, pulse trapping can be explained by an
effective potential that is produced by an accelerating pulse --
the fundamental soliton undergoing SSFS \cite{GorbachSkryabin07}.
The discussion below follows this analysis. The soliton induces a
nonlinear modification of the refractive index, $n$, via the
optical Kerr effect \cite{Agrawal06}:

\begin{equation}
n(I, \omega) = n_0(\omega)+n_2 I, \label{index}
\end{equation}

where $I$ is the intensity and $n_2$ the nonlinear index
coefficient, assuming an instantaneous response, and $n_0$ is the
linear index.

Any light field $A$ in the fiber that interacts with the soliton
sees the nonlinear contribution to the refractive index, $n_2
I=n-n_0$, and experiences cross-phase modulation (XPM)
\cite{PhilbinKuklewicz08}. The evolution of the slowly varying
envelope $A(z, t)$ of this light is governed by \cite{Agrawal06}:

\begin{equation} \frac{\partial A}{\partial z} + \beta_1
\frac{\partial A}{\partial t} + \frac{i \beta_2}{2}
\frac{\partial^2 A}{\partial t^2} = i V A. \label{agpulseprop}
\end{equation}

$z$ and $t$ are space and time in the laboratory frame. $\beta_1$
and $\beta_2$ are the dispersion parameters, where $\beta_n =
\frac{\partial^n}{\partial\omega^n}\big(\frac{n(\omega)
\omega}{c}\big)$ evaluated at the frequency of $A$. Since this
frequency lies in the normal dispersion regime, $\beta_2$ is
positive. $i V A$ is the XPM-term and $V(z, t) = 2 \gamma \vert
A_s(z, t) \vert^2$, where $A_s$ is the envelope of the soliton and
$\gamma$ is the nonlinear parameter.

In the co-moving frame given by $T = t - \beta_1 z$ and $Z=z$,
eqn. (\ref{agpulseprop}) is transformed into the nonlinear
Schr\"odinger equation

\begin{equation}
\frac{\partial A}{\partial Z} + \frac{i \beta_2}{2}
\frac{\partial^2 A}{\partial T^2} = i V A. \label{agpscheqn}
\end{equation}

In analogy to the ordinary Schr\"odinger equation in quantum
mechanics, the XPM contribution $V$ induced by the soliton plays
the part of a potential barrier.

Equation (\ref{agpscheqn}) is transformed into an accelerating
frame \cite{GorbachSkryabin07}:

\begin{equation}
\tau = T + \alpha Z^2/2 \hspace{3cm} \zeta = Z.
\end{equation}

Here $\alpha$ describes the rate of the SSFS, which determines the
soliton acceleration. For a soliton, $\alpha = -8 T_R \beta_{2,
s}^2/15 T_0^4$ \cite{Agrawal06}. $T_R$ is the Raman time; the
soliton length is $T_0$. $\beta_{2, s}$ is the group velocity
dispersion at the soliton wavelength, and the minus sign indicates
that the accelerating soliton is redshifting and slowing down. The
transformation fixes the peak of the accelerating soliton at
$\tau=0$. In this frame, the propagating light $A$ acquires a
phase $\phi(\zeta, \tau)$, so $A(Z, T)$ is replaced with
$\psi(\zeta, \tau)$ \cite{GagnonBelanger90}:

\begin{equation}
\psi = A e^{i\phi} \hspace{3cm} \phi = -\frac{\alpha \zeta
\tau}{\beta_2} + \frac{\alpha^2 \zeta^3}{6 \beta_2}.
\label{subsphase}
\end{equation}

The propagation equation (\ref{agpscheqn}) becomes

\begin{equation}
\frac{\partial \psi}{\partial \zeta} + i \frac{ \beta_2 }{2}
\frac{\partial^2 \psi}{\partial \tau ^2} = i U_{\rm eff}(\tau)
\psi, \label{finaltrappingNLSE}
\end{equation}
where
\begin{equation}
U_{\rm eff}(\tau)=(V(\tau) - \alpha \tau/\beta_2). \label{eqpot}
\end{equation}

\begin{figure}
\begin{center}
\includegraphics[scale=0.3]{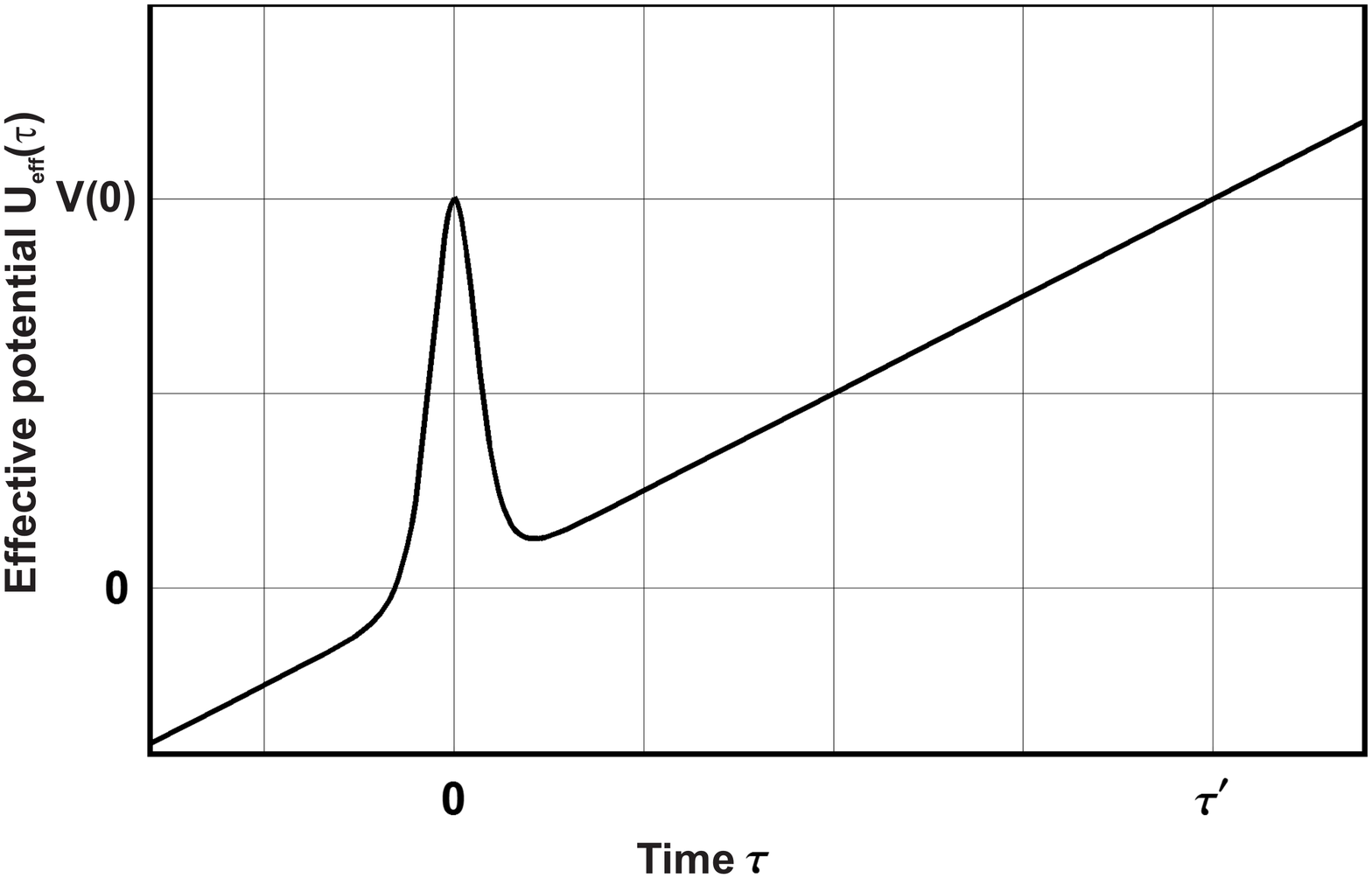}
\caption{The potential created by the accelerating soliton. The
induced nonlinear index change forms a barrier at $\tau = 0$. The
soliton acceleration causes the slope in the potential.
\label{potentialpic}}
\end{center}
\end{figure}

Therefore, in a frame that accelerates with the soliton, other
waves propagate in the presence of an effective potential $U_{\rm
eff}$. The first term in equation \ref{eqpot} is the barrier due
to XPM and the second term is a linear potential due to the
(negative) acceleration $\alpha$. $U_{\rm eff}$ is sketched in
figure \ref{potentialpic}. The barrier is located at $\tau=0$ and
at $\tau = \tau'$ the linear increase reaches the height of the
barrier.

Gorbach and Skryabin pointed out that the potential is
`gravity-like', i.e. similar to the gravitational potential on
earth. Standing in a lift, the floor acts as a strong potential
barrier against the gravitational force. Without gravity, the same
potential would be created if the lift had a constant upward
acceleration. So a uniformly accelerating barrier can trap objects
just as a barrier in a gravitational field does.

As seen in figure \ref{potentialpic}, the potential can trap light
between $\tau=0$ and $\tau=\tau '$. We can estimate the temporal
length of the trap by calculating $\tau'$ using $U_{\rm eff}(\tau
')=U_{\rm eff}(0)$. Approximating $V(\tau')\approx0$ and inserting
$\alpha$ we obtain for an accelerating soliton

\begin{equation}
\tau' = \frac{15 T_0^2}{4 T_R} \frac{\beta_2}{\vert \beta_{2, s}
\vert}. \label{eqtauprime}
\end{equation}

$\tau'$ depends on the SSFS as well as the ratio of group velocity
dispersions for the trapped light and the soliton. In our
experiments the trap length is relatively constant, so that the
trapped light is expected to have a well defined temporal length.
The trap is positioned immediately behind the soliton center at
$\tau\geq0$. In consequence, the trapped light travels at the
group velocity of the soliton and slows down with it. The spectrum
of the trapped light shifts to a wavelength associated with the
same group velocity as the soliton. Accordingly, the potential
determines both the spectral and temporal properties of the
trapped light.

\section{Experiment}

\begin{figure}
\begin{center}
\includegraphics[scale=0.3]{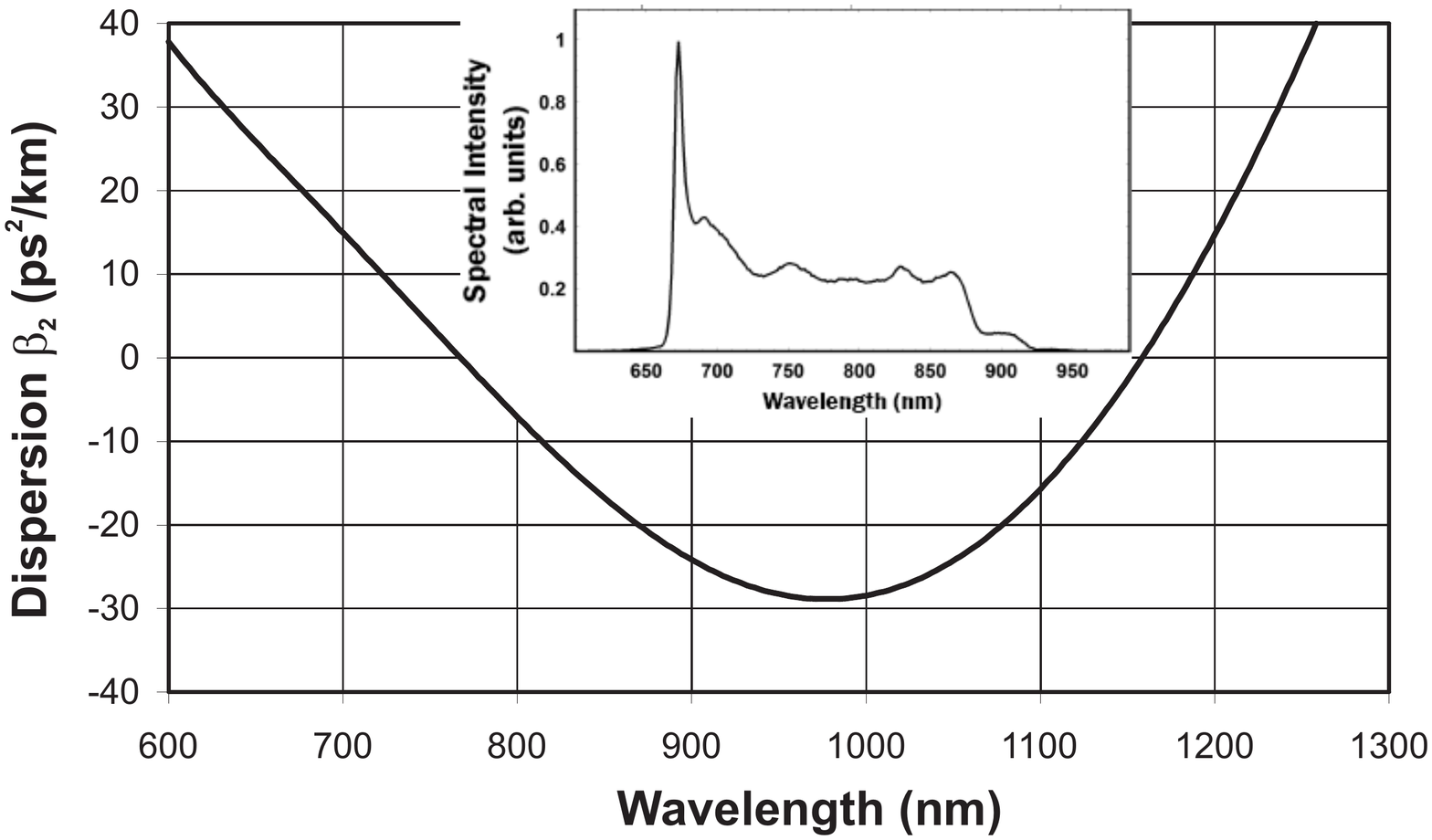}
\caption{The group velocity dispersion, $\beta_2$, of the fiber
NL-PM-760 \cite{crystalfibredata} and the initial spectrum of the
few-cycle pump pulses (inset). The dispersion zeroes at 760\,nm
and 1160\,nm. \label{dispersion and input spec pic}}
\end{center}
\end{figure}

\begin{figure}
\begin{center}
\includegraphics[scale=0.3]{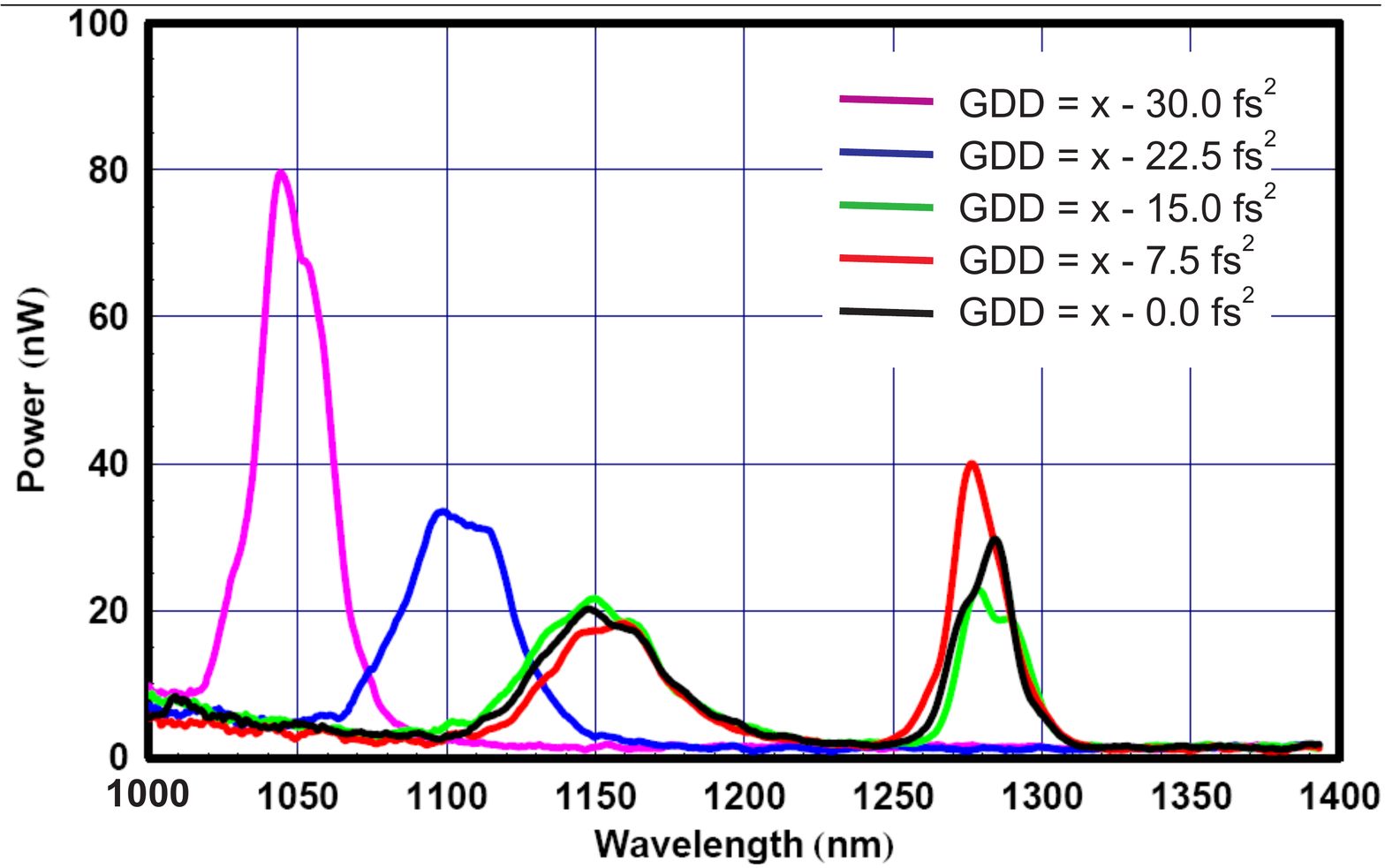}
\caption{Output spectra for varying amounts of chirp on the input
pulse. The Group-Delay-Dispersion (GDD) before the fiber is varied
over 30\,fs$^2$. The pulse energy is 39\,pJ. A GDD of 7.5\,fs$^2$
broadens an unchirped 7-fs pulse by 10\% in time.
\label{varyinginputchirppic}}
\end{center}
\end{figure}

To create the trapping soliton as well as the trapped light we
couple a few-cycle pulse into a microstructured fiber. We use
pulses from a $ < $7\,fs Ti:Sapphire laser with a repetition rate
of 78\,MHz (Rainbow, Femtolasers GmbH). A reflective-optics
telescope expands the beam and a parabolic mirror of high
numerical aperture focuses the light for fiber coupling. The use
of mirrors minimizes dispersion-induced pulse broadening. The
1.8\,m-long fiber (NL-PM-760, Crystal Fiber A/S) has zero
dispersion at 760\,nm and 1160\,nm. The dispersion of the fiber
and the initial pulse spectrum are shown in figure \ref{dispersion
and input spec pic}; the nonlinear coefficient is $\gamma =
102(\rm{W km})^{-1}$ at $\lambda = 780\,$nm
\cite{crystalfibredata}. We rotate the fiber to align the fiber
axes to the polarization of the incoming light. As a test of the
pulse length at the fiber input face we replaced the fiber with a
mirror and backreflected the pulse through the reflective focusing
optics. Compensating for group delay dispersion, no significant
pulse broadening was observed after the double pass. After the
fiber we use a microscope objective to collimate the light and
analyze it temporally with an autocorrelator (Pulsecheck, APE
GmbH). By tilting the nonlinear crystal of the autocorrelator we
perform Frequency-Resolved Optical Gating (FROG) measurements
\cite{Kaertner04}. We also pick off  part of our laser output
beam, reduce the spectral bandwidth with a 60-nm interference
filter, and use it as a reference pulse to record X--FROG traces.
The reference pulse was characterized using a FROG measurement to
be 35-fs in length (FWHM). Simultaneous to the FROG and X-FROG
traces, we measure the spectrum of the trapped light and trapping
solitons on a compact CCD-based spectrometer and an optical
spectrum analyzer (OSA) for longer wavelengths. Figure
\ref{varyinginputchirppic} demonstrates that the fiber output
spectra are very sensitive to the dispersion-induced chirp on the
input pulse. In all experiments we adjusted the input chirp to
obtain maximum output spectral width (black curve in figure
\ref{varyinginputchirppic}).

We create solitons in the near-infrared and trapped light in the
visible. Because the input spectrum lies partially in the
anomalous dispersion regime beyond 760\,nm wavelength, a soliton
forms and Raman shifts via the SSFS. The part of the input light
that is in the normal dispersion regime and that is trailing the
soliton is trapped. In addition there is interaction between light
in the two dispersion regimes via self-phase modulation and
phasematched amplification of dispersive waves
\cite{DudleyProvino02, HerrmannGriebner02, HillingsoAndersen04,
GentyLehtonen04, FroszFalk05}. Therefore, both the trap and
trapped light are created by launching a single pulse into the
fiber. Since the Raman-shifting soliton traps light, the trapped
light has to maintain the soliton group velocity
$1/\beta_1(\lambda_s)$ even though the soliton slows down. The
trapped light impinges onto the soliton which leads to a blueshift
and eventually the maintenance of group velocity matching
\cite{PhilbinKuklewicz08}. Increasing the input pulse energy in
the fiber speeds up the SSFS and secondary fundamental solitons
are created. For high input pulse energies the primary soliton
shifts up to the long ZDW, which it cannot pass. The SSFS abruptly
ends and the soliton decays. As the acceleration of the soliton is
terminated, the `gravity-like' part of the potential disappears,
while the barrier decays slowly, i.e. the trapped light can escape
behind the soliton. At the same time the wavelength shift of the
trapped light ceases.

\subsection{Output Spectra at low pulse energies}

\begin{figure}
\begin{center}
\includegraphics[angle=-90, scale=0.4]{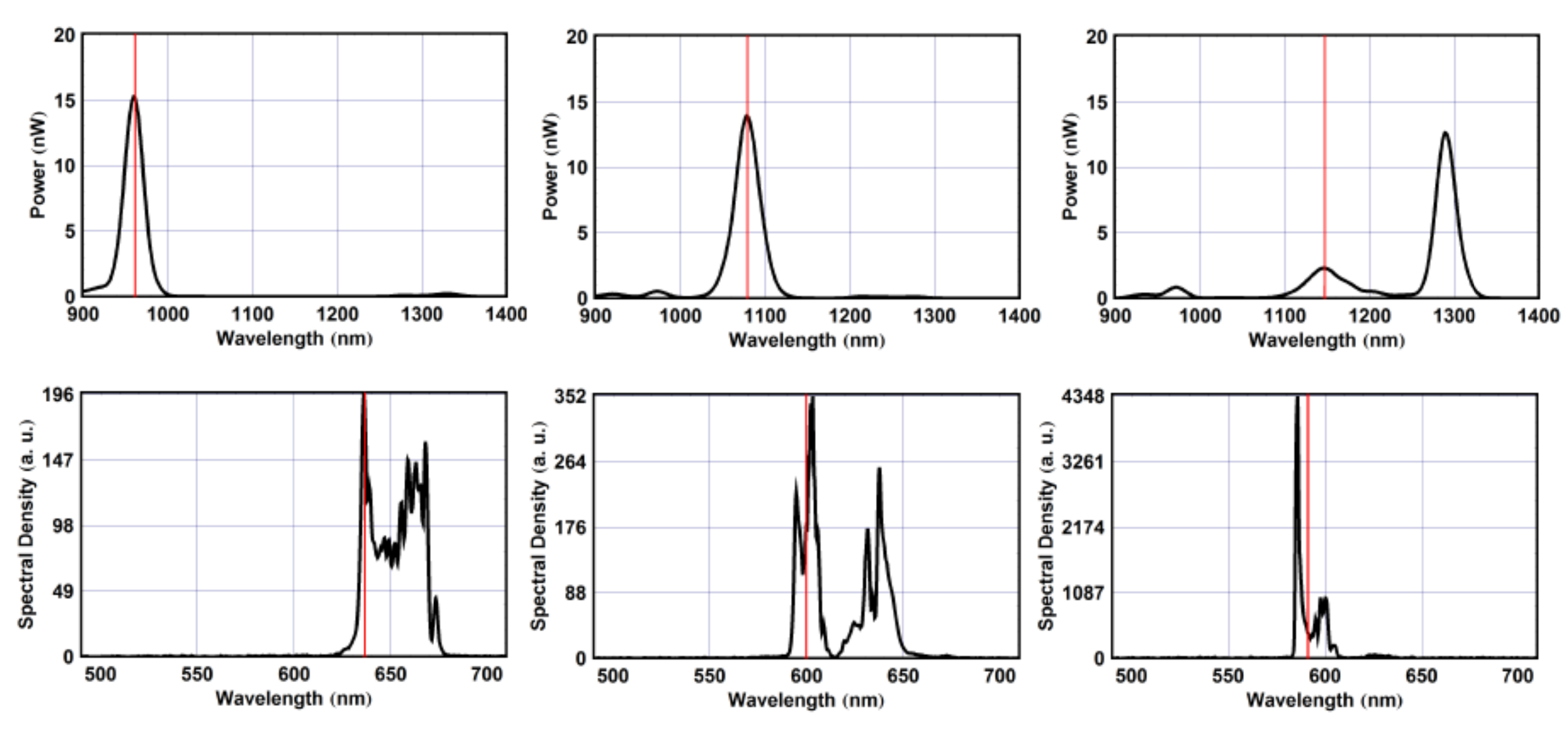}
\caption{Three spectra of the soliton (top) and the trapped light
(bottom) as the launched pulse energy is increased (left: 21\,pJ,
center: 31\,pJ, right: 40\,pJ ). Red lines mark the center
wavelength of the soliton and a group velocity matched wavelength
obtained from data in figure \ref{dispersion and input spec pic}
\cite{crystalfibredata}. \label{lowpowerspecevolution}}
\end{center}
\end{figure}

When sufficient pulse energy is provided, a soliton is generated
and redshifted due to the SSFS. Figure \ref{lowpowerspecevolution}
shows that the rate of shifting increases with pulse energy as
longer wavelengths are reached within our finite fiber length.
Below a wavelength of 700\,nm a trapped pulse is created that
frequency shifts with the soliton, but in opposite direction. The
relative rate of shifting is determined by group velocity
matching. Some of the initial pulse energy remains between 700\,nm
and 900\,nm and propagates ahead of the slowing soliton. Note the
feature near 1300\,nm, which indicates the onset of `Cherenkov'
radiation, as discussed in the next section. Similar spectra have
been observed in other experiments, see for example
\cite{SerebryannikovZheltikov05, FalkFrosz08}.

\subsection{Output spectra at higher pulse energies}

\begin{figure}
\begin{center}
\includegraphics[angle=-90, scale=0.55]{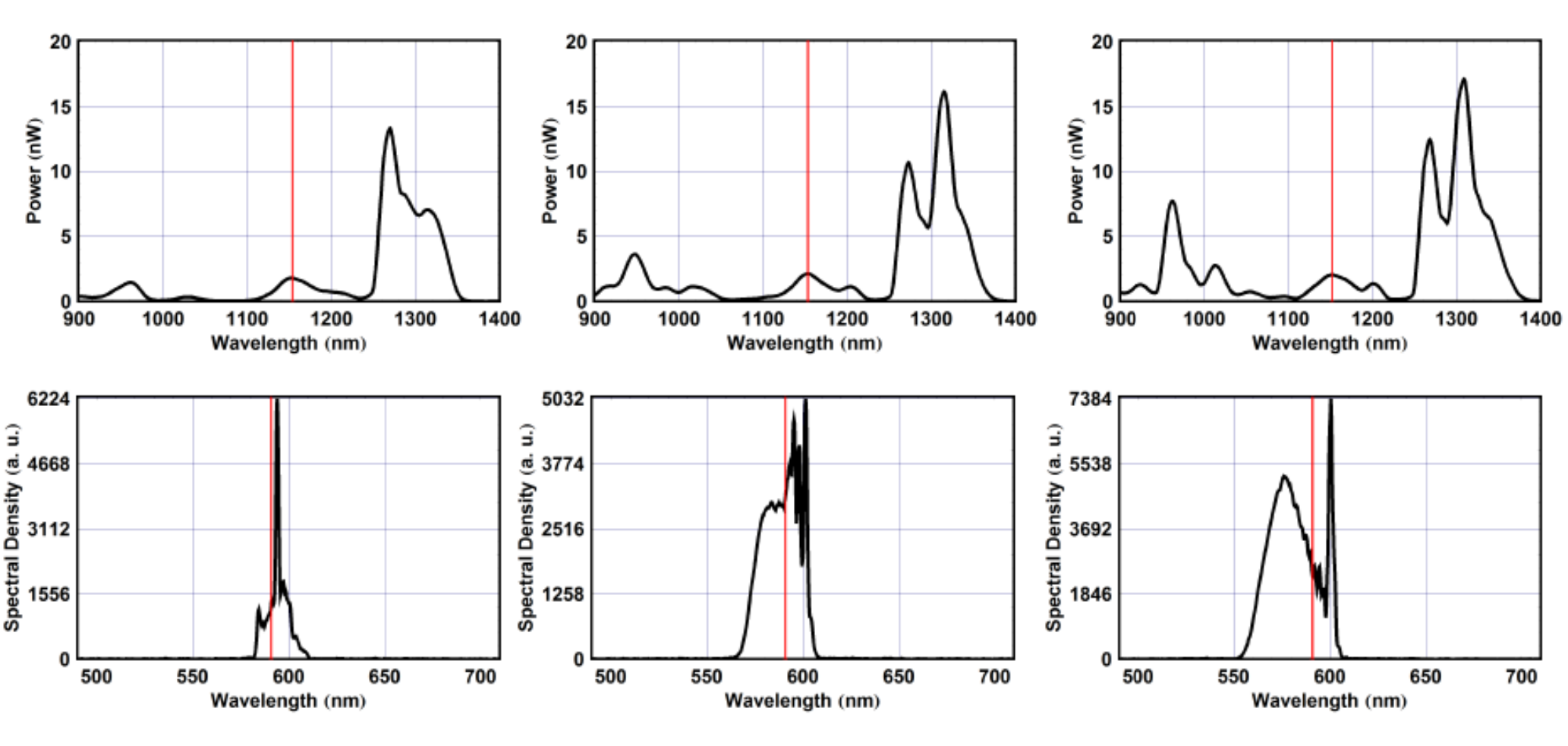}
\caption{Three spectra of the soliton (top) and the trapped light
(bottom) for pulse energies exceeding those of figure
\ref{lowpowerspecevolution} (left: 56\,pJ, center: 88\,pJ, right:
121\,pJ). The Raman shift of the soliton is cancelled by Cherenkov
radiation. The trapped light shifts almost completely beyond the
wavelength that is group velocity matched to the soliton to
shorter wavelengths and trails the soliton. Red line marks as in
figure \ref{lowpowerspecevolution}.
\label{highpowerspecevolution}}
\end{center}
\end{figure}

Spectra for higher pulse energies are shown in figure
\ref{highpowerspecevolution}. The SSFS has ceased and the soliton
has reached a final wavelength of 1150\,nm, just below the ZDW of
1160\,nm. Approaching this wavelength, the soliton produces
dispersive Cherenkov waves around 1300\,nm. The production of
Cherenkov radiation balances the stimulated Raman effect in the
soliton and the soliton redshift terminates. Since the soliton
energy is gradually being transferred to the Cherenkov radiation,
the soliton decays \cite{SkryabinLuan03}.
Figure\ref{highpowerspecevolution} also shows that nearly all of
the trapped light shifts to wavelengths shorter than the group
velocity matched wavelength. These are propagating slower than the
soliton -, that is the light escapes behind the soliton. Light at
wavelengths travelling faster than the soliton collides with the
soliton one last time and shifts to the blue beyond the group
velocity matched wavelength. This is seen particularly clearly by
examining the soliton and the trapped light in the time domain,
which we describe next.

\subsection{FROG and X-FROG measurements at low pulse energies}

\begin{figure}
\hspace{-2.5cm}
\includegraphics[bb= -9cm -59cm 3cm -48cm, scale=0.25]{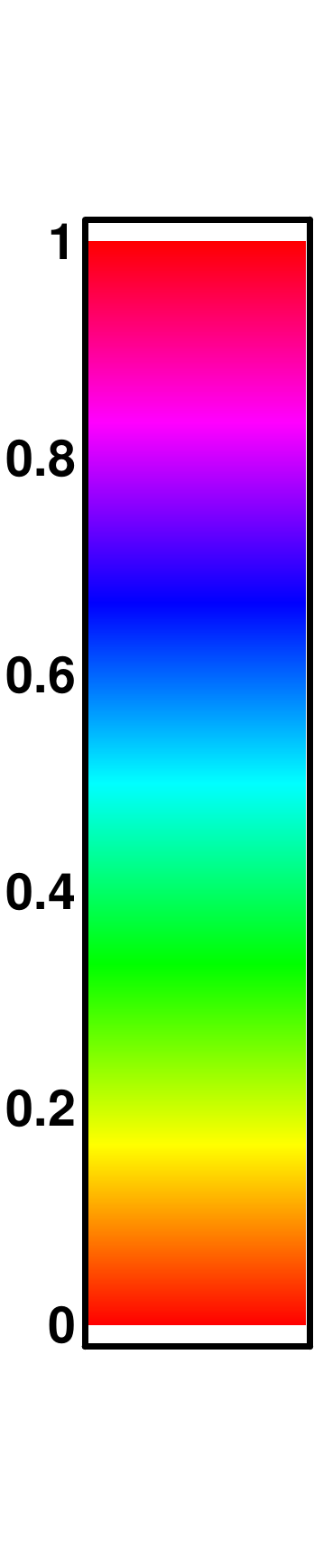}
\includegraphics[scale=0.65]{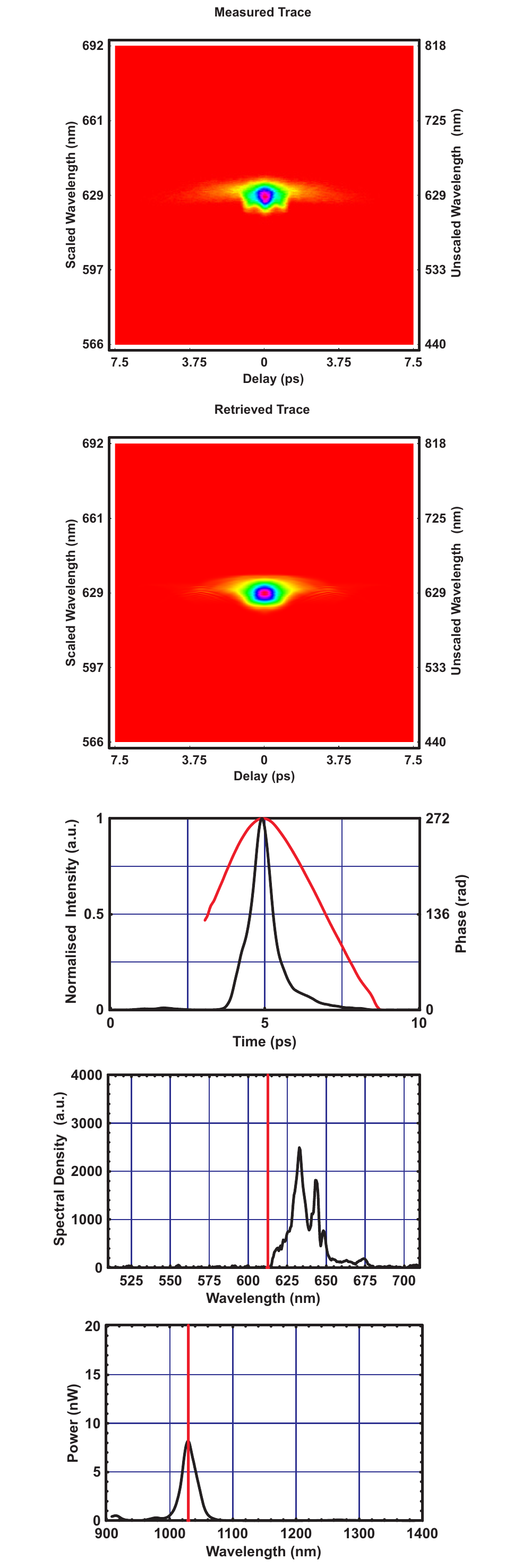}
\includegraphics[scale=0.65]{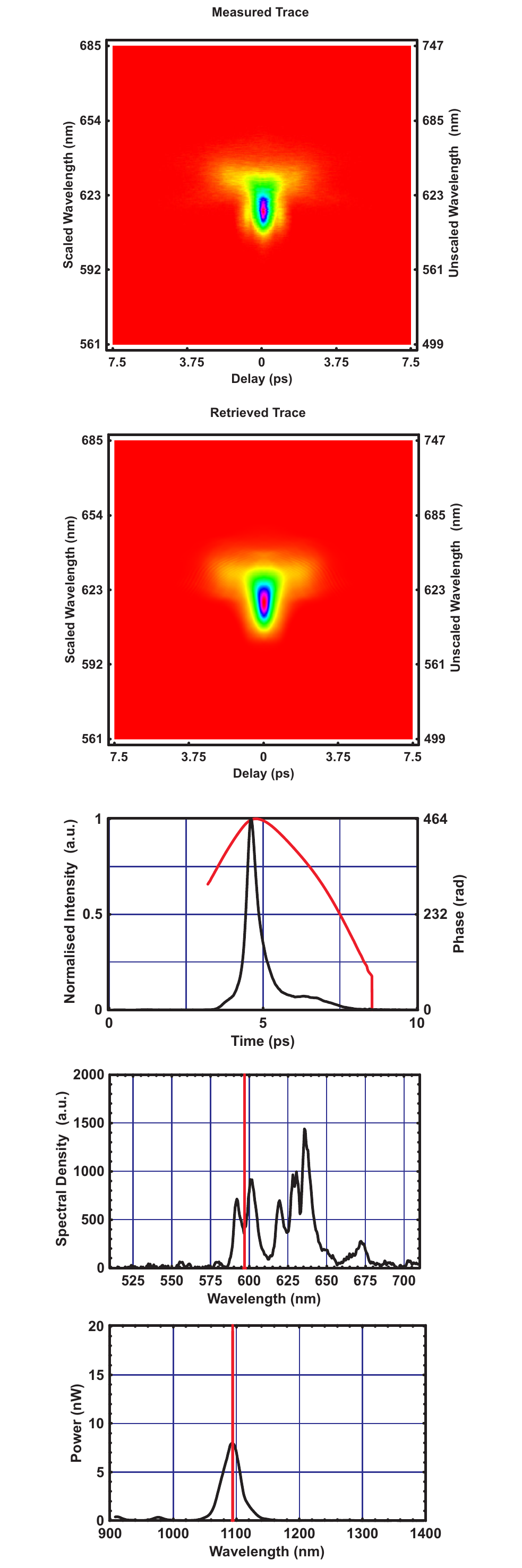}
\caption{Measured SHG-FROG traces (top row), retrieved traces
(second row), and retrieved intensity and phase (third row) for
the trapped light when the soliton is accelerating. Fourth and
fifth row show the independently measured trapped pulse and
soliton spectra, respectively. Pulse energies are: left: 26\,pJ,
right: 32\,pJ. Explanation of FROG axes see text. \label{low power
shg frog}}
\end{figure}

\begin{figure}
\hspace{0.5cm}
\includegraphics[scale=0.16]{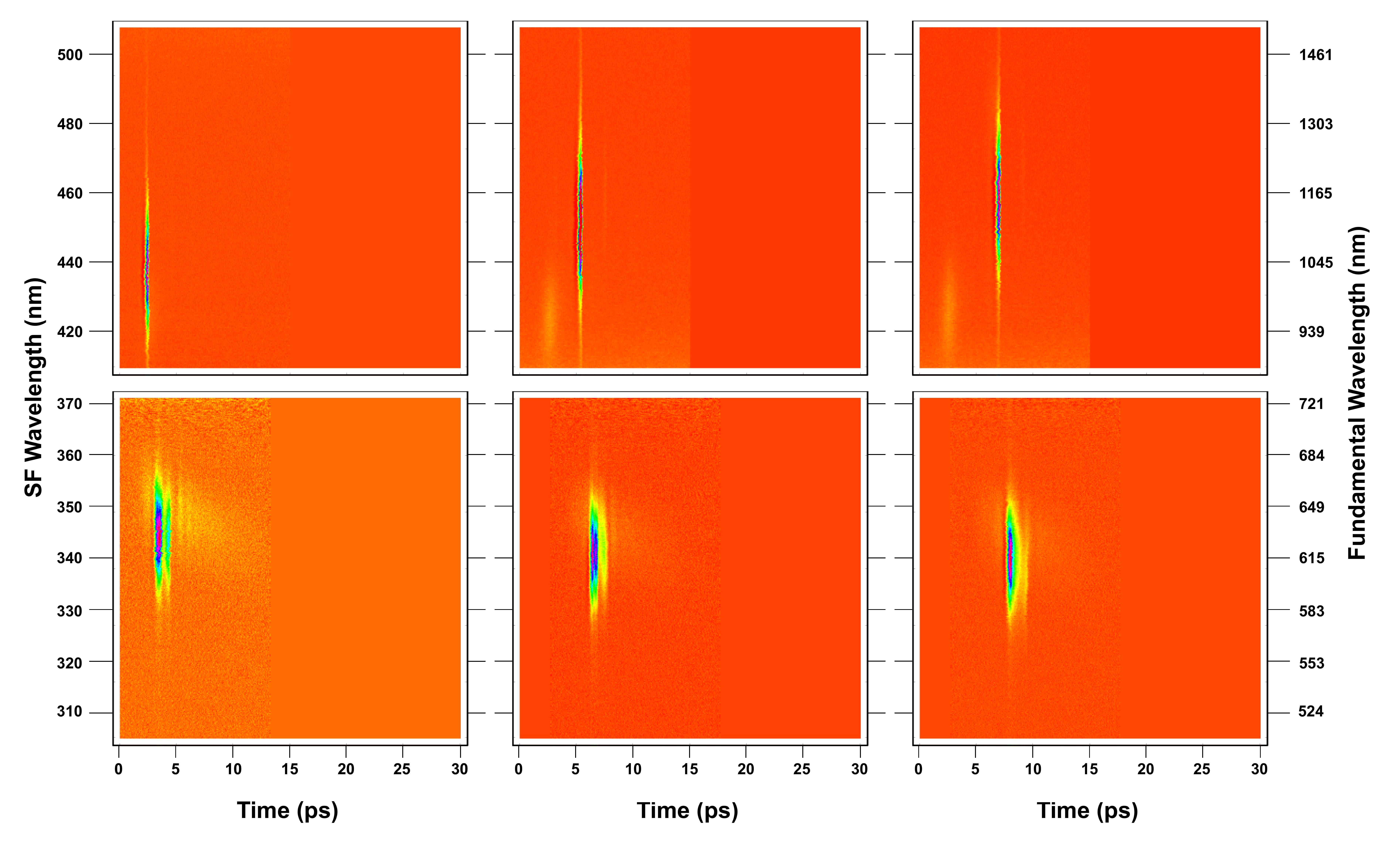}
\caption{X-FROG measurements of the soliton (top) and the
corresponding trapped pulse (bottom) for low input pulse energies.
The direction of propagation is to the left (pulse energies: left:
28\,pJ, center: 34\,pJ, right: 38\,pJ) . Increasing the energy
redshifts (and therefore delays) the soliton. The trapped light is
confined behind the soliton. \label{low power xfrog pic}}
\vspace{-0.5cm}
\end{figure}

For low input pulse energies the soliton undergoes SSFS throughout
the fiber and hence it is always slowing down. The `gravity-like'
part of the trap potential still exists at the fiber end. The
trapped light is confined behind the barrier created by the
soliton. In figure \ref{low power shg frog} second harmonic-FROG
(SHG-FROG) measurements \cite{Kaertner04} of the visible part of
the spectrum show the trapped pulse. All of the FROG and X-FROG
traces in this paper use the same linear color scale shown. The
trapped pulse has a duration of $\sim1$ps. The time-bandwidth
product (FWHM) of this pulse is more than ten -- the spectral
width could support a sub 50-fs pulse.

The FROG retrieval requires a FROG grid as large as 2048x2048
pixels. Before retrieval, the wavelength is scaled by a factor to
fit the data into a square grid \cite{scaling}. After retrieval
the wavelength is unscaled to recover the physical scale. The FROG
error in figure \ref{low power shg frog} is 0.00468 (left column)
and 0.00557 (right column). SHG-FROG has an ambiguity regarding
the direction of time and the FROG measurement is not suitable to
resolve the relative positions in time of the trapped pulse and
soliton. Therefore we performed X-FROG measurements on the
solitons and the trapped pulse by mixing the light from the fiber
with the reference pulse (FWHM 35\,fs) derived before the fiber
and spectrally filtered. The X-FROG traces can be interpreted
directly. Results are shown in figure \ref{low power xfrog pic},
where the propagation direction is to the left. Again the ps-wide
trapped pulse is measured. It is confined to the region behind the
soliton. As the soliton redshifts and delays, the trapped light
blueshifts to maintain group velocity matching with the soliton.
Note that the trapped pulse trails the soliton by about 0.5\,ps.
This delay is due to the normal group velocity dispersion of the
microscope objective used to collimate the light from the fiber.
The measurements clearly show how the trapped pulse is strongly
blocked by the soliton--induced barrier.

The retrieved pulse shapes in figure \ref{low power shg frog}
(left) indicate that the trapped pulse has a small, low intensity
shoulder on one side. The X-FROG measurements in figure \ref{low
power xfrog pic} show that this shoulder extends behind the
trapping soliton and reveal some ps temporal oscillations not
resolved in the retrieved FROG traces. The shoulder seems induced
by the asymmetry of the confining potential. The linearly
increasing part of the trap potential caused by the SSFS is much
more shallow than the sharp edge of the barrier.

\begin{figure}
\hspace{0.0cm}
\includegraphics[scale=0.65]{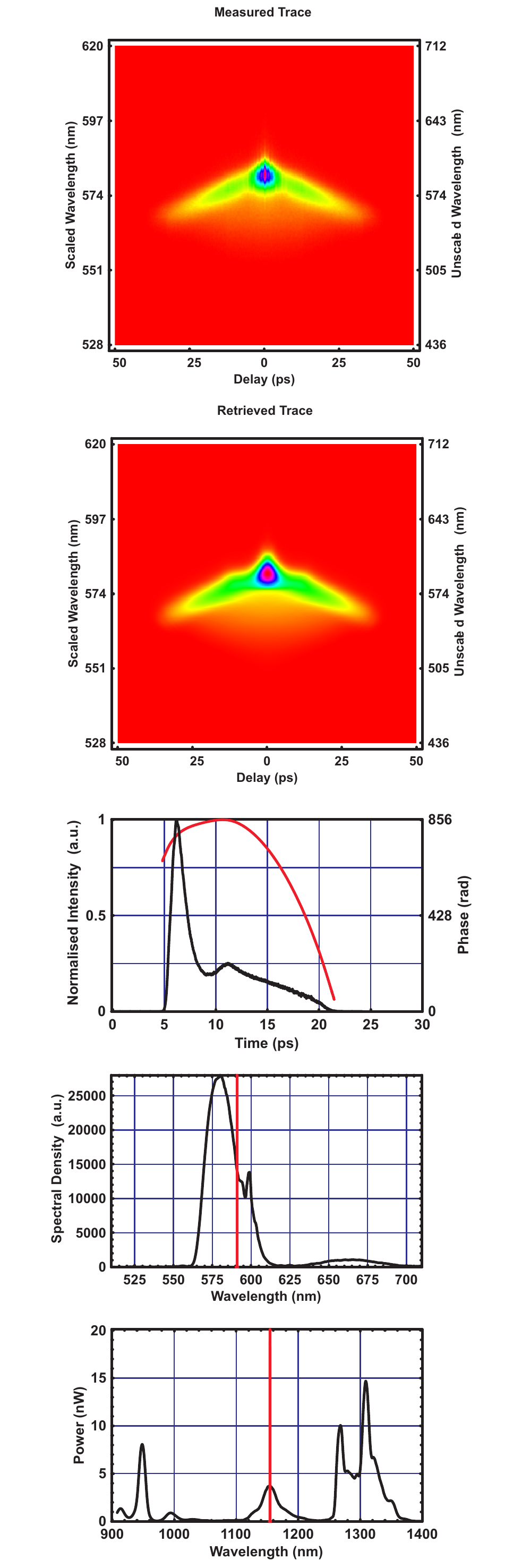}
\includegraphics[scale=0.65]{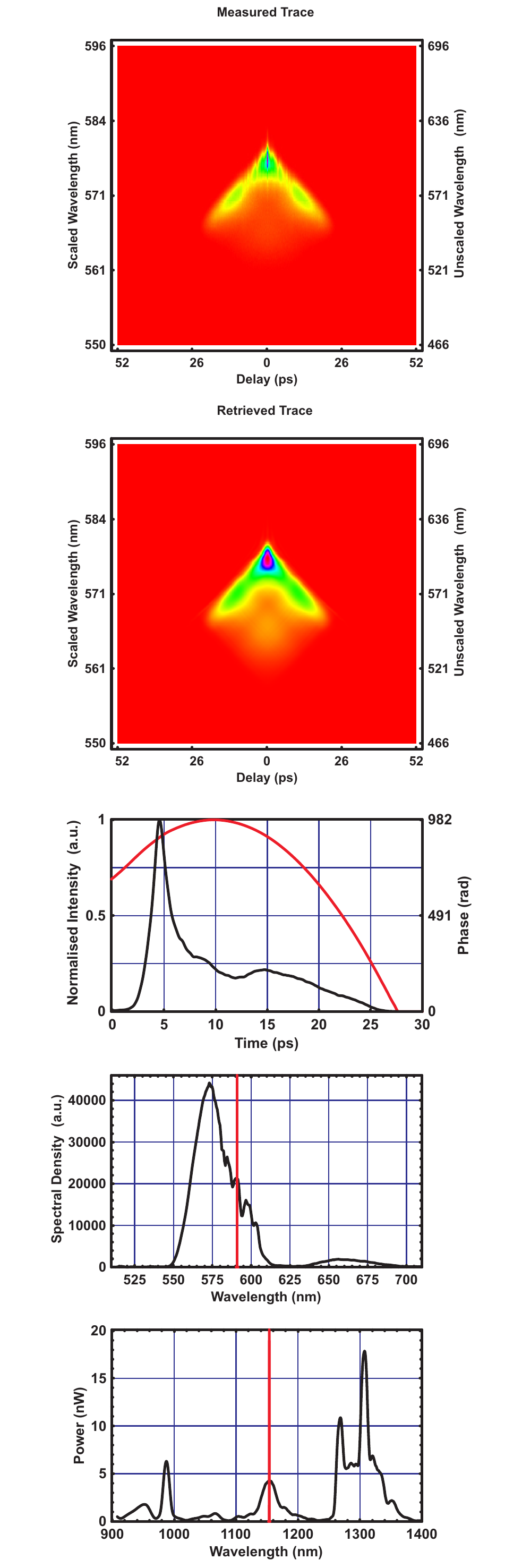}
\caption{Measured SHG-FROG traces (top row), retrieved traces
(second row), and retrieved intensity and phase (third row) for
the trapped light when the soliton has reached the second ZDW and
acceleration is terminated (estimated energies: left: 125\,pJ,
right: 135\,pJ). Rows four and five show the trapped pulse and
soliton spectra, respectively. The trapped pulse has developed a
long tail, which increases beyond the trap length as the trap is
terminated earlier on in the fiber. Explanation of FROG axes see
text.\label{high power shg frog}}
\end{figure}

The size of the trap can be estimated by equation
\ref{eqtauprime}. The soliton, when shifted to a wavelength around
1000\,nm, maintains a near constant width of $T_0 = 60$\,fs for
different input pulse energies as measured separately by
autocorrelation. With $T_R = 5$\,fs \cite{ChengWang05} and a ratio
${\beta_2}/{\vert \beta_{2, s} \vert}\approx 1.5$ (see figure
\ref{dispersion and input spec pic}) this gives a trap length of
4\,ps. The actual length of a pulse in this trap will be somewhat
shorter as the pulse energy will not reach the height of the
potential barrier of the trap due to tunneling. The observed pulse
length of about 1\,ps is in good agreement with this estimate.

\begin{figure}
\hspace{0.5cm}
\includegraphics[scale=0.16]{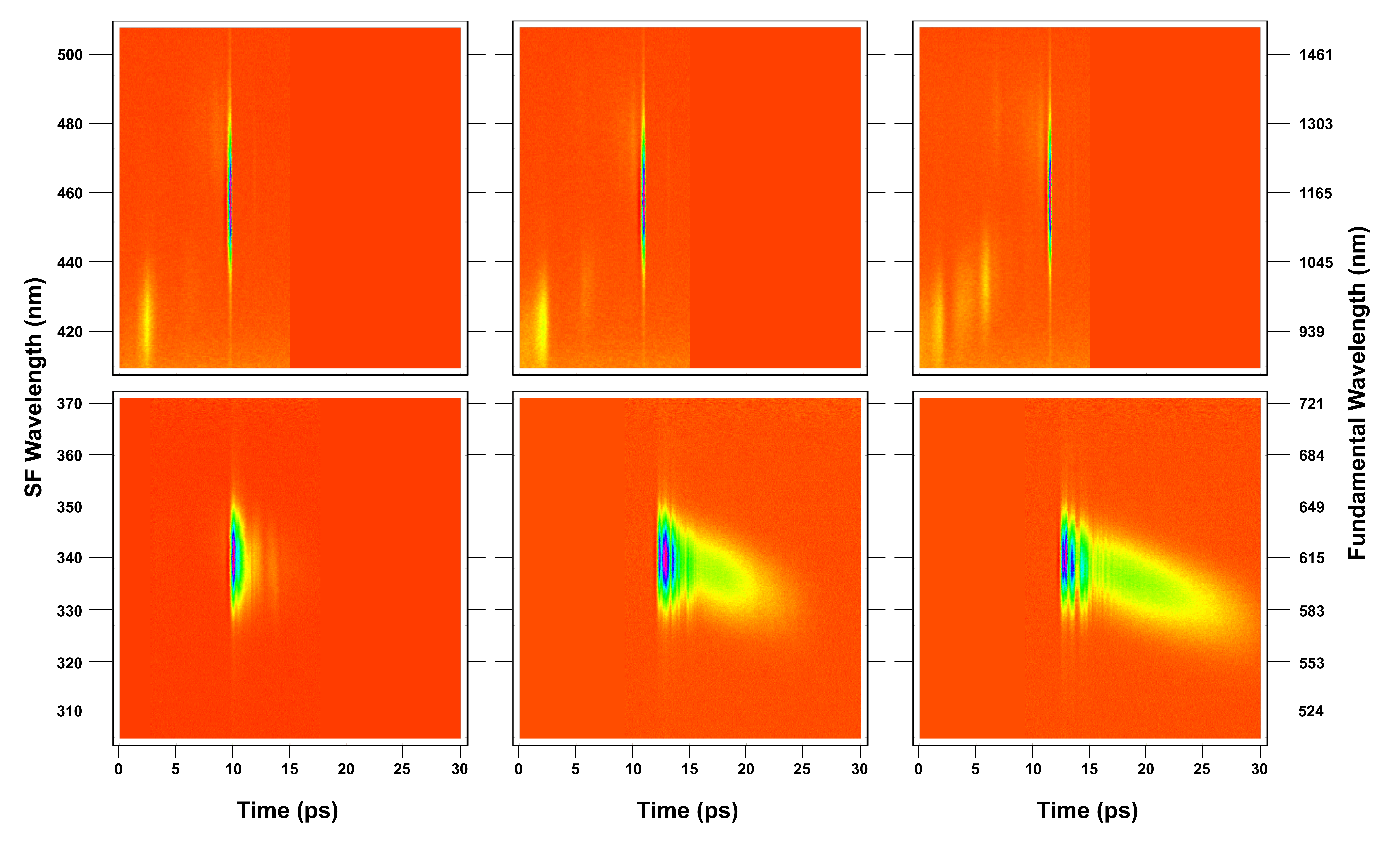}
\caption{X-FROG measurements of the soliton and the trapped pulse
for higher input energies (est. energies: left: 39\,pJ, center:
100\,pJ, right: 120\,pJ). The soliton shift is terminated by the
Cherenkov effect. The trapped light escapes the trap behind the
soliton. The barrier part of the potential is still
present.\label{high power xfrog pic}} \vspace{-0.5cm}
\end{figure}

\subsection{Higher pulse energies and the end of trapping}

When the input pulse energy in the fiber is sufficient for the
soliton to reach the longer zero-dispersion wavelength, the
behavior of the trapped light changes. SHG-FROG measurements of
the trapped light are shown in figure \ref{high power shg frog}.
(FROG retrieval errors: 0.00449 (left), 0.00484 (right)). The
pulse peak has widened considerably beyond 1\,ps and the small
shoulder of figure \ref{low power shg frog} has developed into a
long tail. With more energetic input pulses the tail extends
further.

This behavior again is a direct consequence of the soliton
dynamics. When the soliton wavelength is stabilized by Cherenkov
radiation, the acceleration terminates rapidly, but the soliton
decays slowly \cite{SkryabinLuan03}. The barrier part of the
potential remains, but the gravity-like part disappears. Hence one
side of the potential well vanishes and the light escapes the trap
to one side creating the pronounced tail in figure \ref{high power
shg frog}. A higher launched pulse energy leads to a faster SSFS
and the soliton acceleration turns off earlier on in the fiber.
Thus the trapping ceases earlier and the trapped light can spread
out more under dispersion during the remaining length of fiber.
This is seen in figure \ref{high power shg frog} as the tail
extends further for higher pulse energies.

Corresponding X-FROG traces of the trapped pulse and soliton after
the soliton redshift has ceased are shown in figure \ref{high
power xfrog pic}. The soliton remains a barrier and prevents the
trapped pulse from overtaking as before. On the other side the
developing tail is clearly visible, extending more than 10\,ps
behind the soliton barrier (see also figure \ref{high power shg
frog}). Again the tail bears oscillations not resolved in the
retrieved FROG traces. The shorter wavelengths in the tail travel
slower due to the normal dispersion. The spectrum of the trapped
pulse in figure \ref{high power shg frog} (right) is centered at
580\,nm with a spectral width of 40\,nm. At this wavelength the
fiber dispersion is $\beta_2\approx45$ ps$^2$/km (figure
\ref{dispersion and input spec pic}).  This leads to an expected
length of the tail of about 10\,ps per meter, in good agreement
with the measured length. Thus as soon as the trapping terminates,
the trapped light spreads quickly.

\section{Yet higher pulse energies}

\begin{figure}
\hspace{0.5cm}
\includegraphics[angle=-90, scale=0.4]{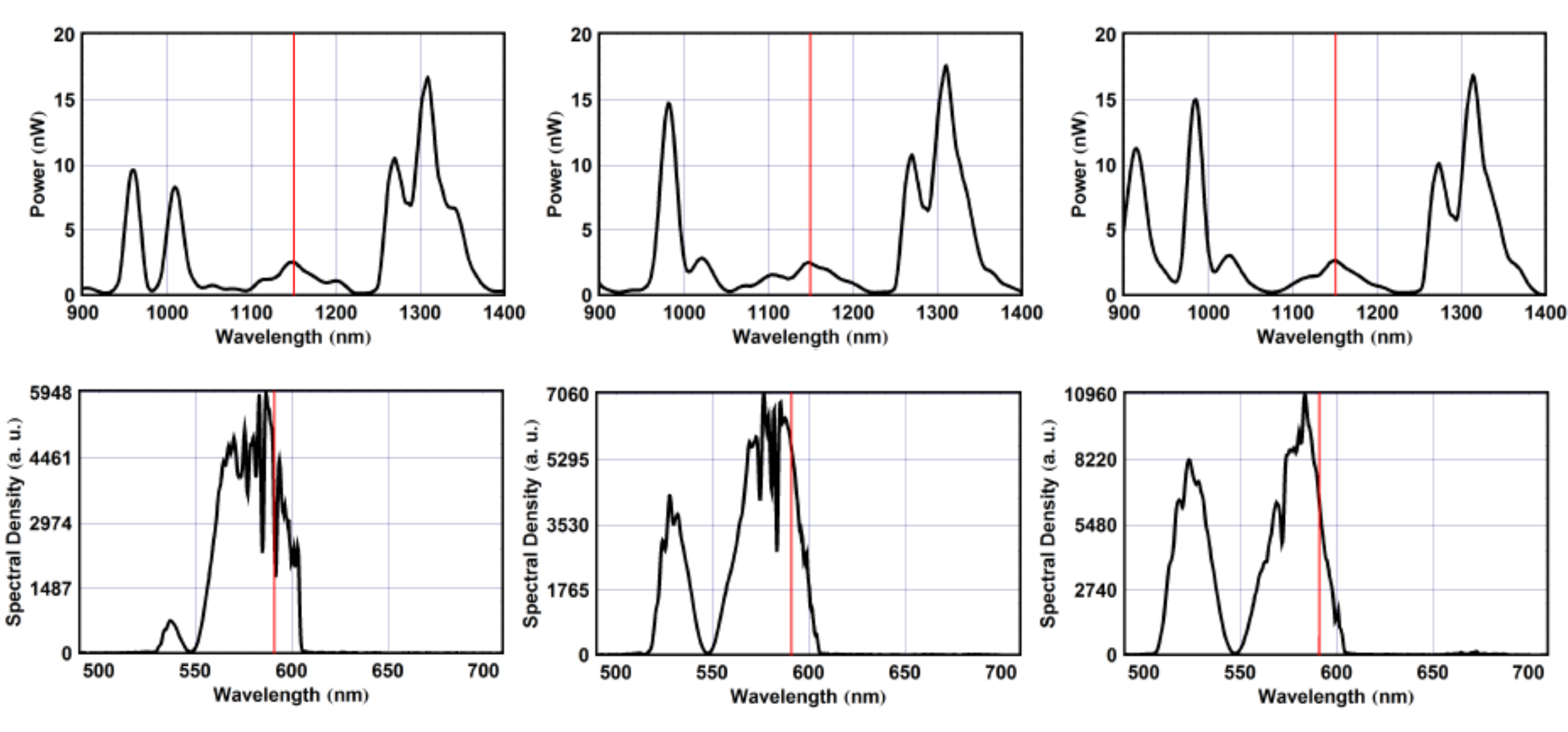}
\caption{The spectrum of the soliton (top) and the
short-wavelength end of the supercontinuum (bottom) for pulse
energies exceeding those of figure \ref{highpowerspecevolution}
(left: 170\,pJ, center: 210\,pJ, right: 280\,pJ) . The soliton
reaches the longer ZDW even earlier in the fiber and multiple
solitons undergo SSFS (top). The trapped light is released, but
the spectrum extends further to the blue with a distinctive gap at
550\,nm (bottom). \label{veryhighpowerspecevolution}}
\end{figure}

\begin{figure}
\begin{center}
\includegraphics[scale=0.27]{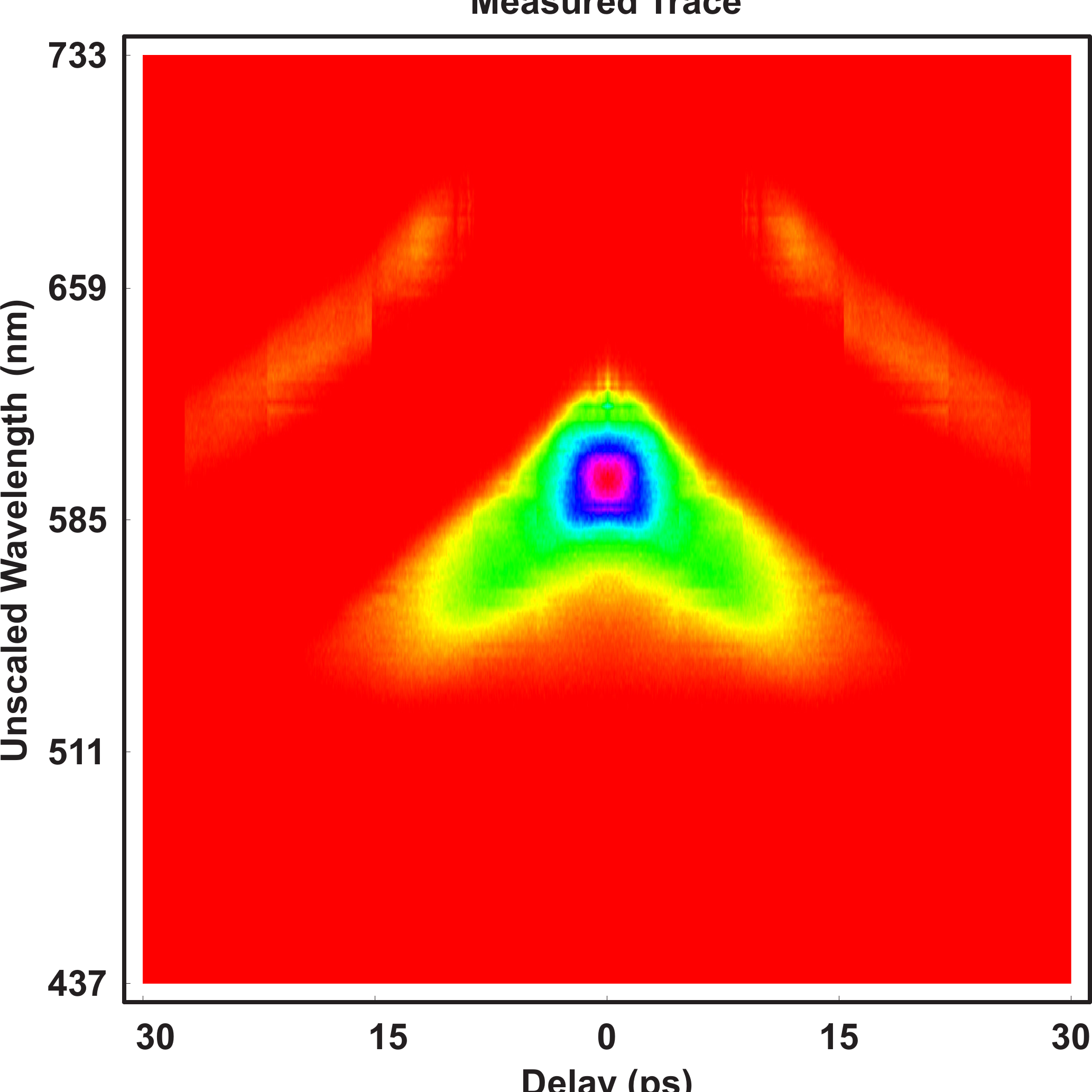}
\caption{SHG-FROG measurement of the trapped pulse for a pulse
energy of 295\,pJ corresponding to figure
\ref{veryhighpowerspecevolution}. The trace is similar to previous
FROG measurements of the trapped light after the trapping has
ended (see figure \ref{high power shg frog}).
\label{veryhighpowerSHG}}
\end{center}
\end{figure}

By sacrificing the reference beam used in the X--FROG measurements
we are able to further increase the pulse energy into the fiber.
The soliton now reaches the wavelength stabilized by Cherenkov
radiation even earlier in the fiber. Unfortunately, the large
temporal width of the pulse and a small signal appearing at longer
wavelengths made it impossible for our code to retrieve the FROG
trace. The FROG trace is similar to traces of the trapped light at
lower powers (compare to figure \ref{high power shg frog}). The
length of the tail is determined by the dispersion over the full
fiber length. The wavelength of the pulse peak is arrested at
$\sim$590\,nm. This stabilizes the trapped pulse and makes it
independent of coupled input energy in this energy range.

The spectrum continues to reach shorter wavelengths, considerably
below the group velocity-matched point (figure
\ref{veryhighpowerspecevolution}). We attribute the additional
peak of light appearing below 550\,nm to nonlinearly phase matched
resonant wave mixing at the very input of the fiber
\cite{YulinSkryabin04, DudleyProvino02, TseHorak06}. This is
consistent with the observation that the peak shifts to the blue
with increasing energy. Similar behaviour has been seen in
\cite{HillingsoAndersen04,FalkFrosz08}. In our case there is a
particularly pronounced spectral gap at 550\,nm, independent of
input energy and of constant shape. In the infrared we see that
multiple solitons, created from the input pulse by soliton
fission, have shifted towards the long ZDW. The soliton that is
arrested just short of the long ZDW is at a minimum of group
velocity. The initially trapped light still trails the soliton and
thus is of a wavelength shorter than the group velocity-matched
wavelength. Light in the trap that had been propagating $\Delta v$
faster than the soliton before release has undergone a spectral
blueshift at the soliton to a group velocity $\Delta v$ slower
than the soliton \cite{PhilbinKuklewicz08}.

\section{Conclusion}

By measurements in both the temporal and spectral domains, we show
how the dynamics of a soliton determine both the wavelength and
pulse-like nature of the short wavelength end of the
supercontinuum. Light in the normal dispersion regime forms a
trapped pulse governed by a potential consisting of a nonlinear
barrier and a `gravity-like' linear potential. The linear part is
due to the SSFS which causes a negative acceleration of the
soliton. As long as the soliton is accelerating, the trapping
confines the light in time. When the SSFS terminates, the trapped
light falls behind the soliton, moving to a wavelength associated
with a group--velocity slower than the soliton. This effect
dominates the shape of the supercontinuum at short wavelengths
until resonant wave mixing with the input pulse further broadens
the spectrum.

This setup allows for the creation and release of wavelength
tunable picosecond pulses in the normal dispersion regime at any
location along a microstructured fiber. The wavelength for the
trapped pulse is obtained by choosing an appropriate dispersion
profile of the fiber. The most important parameter is the longer
zero dispersion wavelength. If the fiber has a long ZDW that is
longer than 1160\,nm, then the SSFS would take the soliton further
into the infrared before the acceleration terminates. The trapped
pulse would shift to even shorter wavelengths and still be
confined in time. Once the soliton is arrested, the trapped pulse
has reached the final wavelength which is largely independent of
pulse energy. Hence, the trapped pulse wavelength could be
increased from this by use of lower input pulse energies or
chirped input pulses.

\section{Acknowledgements}

We are indebted to Dmitry Skryabin, Andrey Gorbach, Scott
Robertson, Franz K\"artner, Peter Staudt, Klaus Metzger and Wilson
Sibbett for discussions and technical support. This work is
supported by the EPSRC, the Royal Society, and the Leonhardt Group
Aue.


\begin{thebibliography}{99}

\bibitem{DudleyGenty06} J. M. Dudley, G. Genty, S. Coen,
``Supercontinuum Generation in Photonic Crystal Fiber,'' \rmp {\bf
78,} 1135--1184 (2006).

\bibitem{IshiiTeisset06} N. Ishii, C. Y. Teisset, S. Kohler, E. E.
Serebryannikov, T. Fuji, T. Metzger, F. Krausz, A. Baltuska, A. M.
Zheltikov, ``Widely Tunable Soliton Frequency Shifting of
Few-Cycle Laser Pulses,'' \pre {\bf 74,} 036617 (2006).

\bibitem{SerebryannikovZheltikov05} E. E. Serebryannikov, A. M.
Zheltikov, N. Ishii, C. Y. Teisset, S. Kohler, T. Fuji, T.
Metzger, F. Krausz, A. Baltuska, ``Soliton Self-Frequency Shift of
6-fs Pulses in Photonic-Crystal Fibers,'' \apb {\bf 81,} 585--588
(2005).

\bibitem{FalkFrosz08} P. Falk, M. H. Frosz, O. Bang, L. Thrane,
P. E. Andersen, A. O. Bjarklev, K. P. Hansen, and J. Broeng,
``Broadband light generation at $\sim$ 1300nm through spectrally
recoiled solitons and dispersive wave ,'' \ol {\bf 33,} 621--623
(2008).

\bibitem{NishizawaGoto02-1} N. Nishizawa and T. Goto,
``Characteristics of pulse trapping by use of ultrashort soliton
pulses in optical fibers across the zero-dispersion wavelength ,''
\opex {\bf 10,} 1151--1159 (2002).

\bibitem{NishizawaGoto02-2} N. Nishizawa, T. Goto ``Pulse Trapping
by Ultrashort Soliton Pulses in Optical Fibers Across
Zero--Dispersion Wavelength,'' \ol {\bf 27,} 152--154 (2002).

\bibitem{NishizawaGoto03} N. Nishizawa, T. Goto ``Ultrafast All
Optical Switching by Use of Pulse Trapping Across Zero--Dispersion
Wavelength,'' \opex {\bf 11,} 359--365 (2003).

\bibitem{GorbachSkryabin07} A. V. Gorbach, D. V. Skryabin ``Light
Trapping in Gravity--Like Potentials and Expansion of
Supercontinuum Spectra in Photonic--Crystal Fibres,'' Nature
Photonics {\bf 1,} 653--656 (2007).

\bibitem{TraversRulkov08} J. C. Travers, A. B. Rulkov, B. A.
Cumberland, S. V. Popov, and J. R. Taylor ``Visible supercontinuum
generation in photonic crystal fibers with a 400\,W continuous
wave fiber laser,'' \opex {\bf 16,} 14435--14447 (2008).

\bibitem{CumberlandTravers08} B. A. Cumberland, J. C. Travers, S.
V. Popov, and J. R. Taylor ``Toward visible cw-pumped
supercontinua,'' \ol {\bf 33,} 2122--2124 (2008).

\bibitem{HusakouHerrmann01} A. V. Husakou and J. Herrmann,
``Supercontinuum generation of higher-order solitons by fission in
photonic crystal fibers ,'' \prl {\bf 87,} 203901 (2001).

\bibitem{Agrawal06} G. P. Agrawal, {\it Nonlinear Fiber Optics}
(Academic Press, 2006).

\bibitem{PhilbinKuklewicz08} T. G. Philbin, C. Kuklewicz, S.
Robertson, S. Hill, F. K\"onig, U. Leonhardt, ``Fiber--Optical
Analog of The Event Horizon,'' Science {\bf 319,} 1367--1370
(2008).

\bibitem{GagnonBelanger90} L. Gagnon, P. A. B\'elanger, ``Soliton
Self--Frequency Shift Versus Galilean--Like Symmetry,'' \ol {\bf
15,} 466--468 (1990).

\bibitem{crystalfibredata} Measured data provided by Crystal Fibre
A/S, Denmark.

\bibitem{Kaertner04} F. X. K\"artner, {\it Few-Cycle Laser Pulse
Generation and Its Applications: Vol 95} (Springer, 2004).

\bibitem{DudleyProvino02} J. M. Dudley, L. Provino, N. Grossard, H.
Mailotte, R. S. Windeler, B. J. Eggleton, S. Coen ``Supercontinuum
generation in air-silica microstructured fibers with nanosecond
and femtosecond pulse pumping,'' \josab {\bf 19,} 765--771 (2002).

\bibitem{HerrmannGriebner02} J. Herrmann, U. Griebner, N.
Zhavoronkov, A. Husakou, D. Nickel, J. C. Knight, W. J. Wadsworth,
P. St. J. Russell and G. Korn, ``Experimental Evidence for
Supercontinuum Generation by Fission of Higher-Order Solitons in
Photonic Fibers,'' \prl {\bf 88,} 173901 (2002).

\bibitem{HillingsoAndersen04} K. M. Hilligs\o e, T. V. Andersen, H.
N. Paulsen, C. K. Nielsen, K. M\o lmer, S. Keiding, R.
Kristiansen, K. P. Hansen, and J. J. Larson, ``Supercontinuum
generation in a photonic crystal fiber with two zero dispersion
wavelengths ,'' \opex {\bf 12,} 1045--1054 (2004).

\bibitem{GentyLehtonen04} G. Genty, M. Lehtonen, H. Ludvigsen, and
M. Kaivola ``Enhanced bandwidth of supercontinuum generated in
microstructured fibers ,'' \opex {\bf 12,} 3471--3480 (2004).

\bibitem{FroszFalk05} M. H. Frosz, P. Falk, and O. Bang ``The role
of the second zero-dispersion wavelength in generation of
 supercontinua and bright-bright soliton-pairs across the
 zero-dispersion wavelength ,'' \opex {\bf 13,} 6181--6192 (2005).

\bibitem{ChengWang05} C. Cheng, X. Wang, Z. Fang, B. Shen,
``Nonlinear copropagation of two optical pulses of different
frequencies in photonic crystal fiber,'' \apb {\bf 80,} 291--294
(2005).

\bibitem{SkryabinLuan03} D. V. Skryabin, F. Luan, J. C. Knight, P.
St. J. Russell, ``Soliton Self--Frequency Shift Cancellation in
Photonic Crystal Fibers,'' Science {\bf 301,} 1705--1708 (2003).

\bibitem{scaling} The applied scaling factor was as follows: Fig.
\ref{low power shg frog} left:3, right:2; Fig. \ref{high power shg
frog} left:3, right:2.

\bibitem{YulinSkryabin04} A. V. Yulin, D. V. Skryabin, and
P. St. Russell ``Four-wave mixing of linear waves and solitons in
fibers with higher-order dispersion,'' \ol {\bf 29,} 2411--2413
(2004).

\bibitem{TseHorak06} M. L. V. Tse, P. Horak, F. Poletti, N. G. R.
Broderick, J. H. V. Price, J. R. Hayes, and D. J. Richardson
``Supercontinuum generation at 1.06$\mu$ m in holey fibers with
dispersion flattened profiles,'' \opex {\bf 14,} 4445--4451
(2006).

\end{thebibliography}
\end{document}